\documentclass{PoS}

\usepackage{amssymb,amsmath}
\newcommand{\csw}{c_{sw}}
\newcommand{\Dlr}{\overset{\leftrightarrow}{D}}
\newcommand{\gR}{g_{\mathrm R}}
\newcommand{\cO}{\mathcal {O}}
\newcommand{\MS}{{\overline{\mbox{MS}}}}
\newcommand{\preprintline}{\newline
\vspace{-4.3cm}
\rightline{\parbox{12cm}
            {\hfill \rm\small DESY 06-181, Edinburgh 2006/26, Leipzig LU-ITP 2006/15, Liverpool LTH 719}}
\vspace{2.0cm}}

\title{One-loop Renormalisation of Lattice QCD Operators for Non-forward
Matrix Elements: From Clover to Overlap Fermions\preprintline}

\ShortTitle{One-loop Renormalisation of Lattice QCD Operators for
Non-forward Matrix Elements}

\author{QCDSF collaboration:
   M.~G\"ockeler,$^a$ R.~Horsley,$^b$ H.~Perlt,$^c$
   P.~E.~L.~Rakow,$^d$ G.~Schierholz,$^e$
   A.~Schiller$^{c,}$\thanks{Speaker (Arwed.Schiller@itp.uni-leipzig.de).} 
\\   
\llap{$^a$}Institut f\"ur Theoretische Physik, Universit\"at Regensburg, 
  93040 Regensburg, Germany\\
\llap{$^b$}School of Physics, University of Edinburgh,
Edinburgh EH9 3JZ, UK\\
\llap{$^c$}Institut f\"ur Theoretische Physik, Universit\"at 
           Leipzig, 04109 Leipzig, Germany\\
\llap{$^d$}Theoretical Physics Division, Department of Mathematical Sciences,
University of Liverpool, Liverpool L69 3BX, UK\\
\llap{$^e$}John von Neumann Institut NIC/DESY Zeuthen, 15738 Zeuthen Germany, \\
Deutsches Elektronen-Synchrotron DESY, 22603 Hamburg, Germany}

\abstract{We consider the renormalisation of composite quark-antiquark
operators with one and two lattice covariant derivatives related
to the lowest moments of generalised parton distributions  (GPDs)
and meson distribution amplitudes (DAs).
Their  matrix elements are calculated in one-loop
lattice perturbation theory for non-zero momentum transfer.
Using clover and overlap fermions we present 
the resulting matrices of mixing and renormalisation factors.
For overlap fermions we explicitly check the absence of
mixing with lower-dimensional operators of different chirality
in  particular representations of the hypercubic group.
This feature favours the use of chiral fermions.}

\FullConference{XXIVth International Symposium on Lattice Field Theory\\
                July 23-28, 2006\\
                Tucson, Arizona, USA}

\begin{document}

\section{Introduction}

Many interesting observables in hadron physics, e.g.\ (moments of)
generalised parton distributions (GPDs) or distribution amplitudes (DAs),
can be computed from matrix elements of local operators between hadron states.
(For an extensive review of GPDs see Ref.~\cite{Diehl:2003ny}, for DAs 
see, e.g., Ref.~\cite{DA}.)
GPDs parametrise a
large class of hadronic correlators, including  form factors
and the ordinary parton distribution functions. Thus those distributions provide
a formal framework to connect information from various inclusive,
semi-inclusive and exclusive reactions.
Furthermore they give access to physical quantities which
cannot be directly determined in experiments, like e.g.\ the orbital
angular momentum of quarks and gluons in a nucleon (in a given
scheme) and the spatial distribution of the energy or spin
density of a fast moving hadron in the transverse plane.
Since the structure of GPDs is rather complicated 
direct experimental
access is limited. Therefore, complementary
information from lattice QCD should be used.
For a recent overview on Monte Carlo results, see Ref. ~\cite{Orginos}.

To make contact with the continuum, the lattice operators needed for GPDs and DAs
have to be renormalised.
Compared with moments of ordinary parton distributions, the specific
difficulty in the treatment of moments of GPDs and DAs 
lies in the fact that the required matrix elements are no longer forward matrix elements.
This circumstance complicates the pattern of mixing under
renormalisation.

In Ref.~\cite{Gockeler:2004xb} we have calculated the
non-forward quark matrix elements needed for the renormalisation of
quark-antiquark operators with two derivatives, which determine
the second moments of GPDs, and we have discussed the mixing problem
in detail. This calculation was performed in one-loop lattice
perturbation theory for the Wilson fermion action.

Here we present some of our results~\cite{Gockeler:2006nb} obtained
for $O(a)$ improved fermions using the Sheikholesla\-mi--Wohlert (clover)
action~\cite{Sheikholeslami:1985ij} (without operator improvement) and new 
calculations for overlap fermions~\cite{Neuberger:1997fp}.

The fermion action  is of the following generic form (spinor and colour indices 
are suppressed)
\begin{equation}
  S_F= \bar{\psi} D \psi \equiv a^4 \sum_{x,y} \bar{\psi}(x) D(x-y) \psi(y)
\end{equation}
with
\begin{eqnarray}
  D_{SW} &=& D_W - {\rm i} \,  a \, g \, r \ \csw \frac{1}{8} [\gamma_\mu , \gamma_\nu ] \,
  F^{\rm clover}_{\mu\nu} \,,
  \\
  D_N&=&= \frac{\rho}{a} \left( 1 + \frac{X}{\sqrt{X^\dagger X}} \right) \, , 
  \quad X= D_W -\frac{\rho}{a}
\end{eqnarray}
for the clover ($D_{SW}$) and overlap ($D_N$) operator with massless quarks, respectively.
Here $D_W$ denotes the Wilson-Dirac fermion operator with the forward (backward)
covariant derivative $\Delta_\mu$ ($\Delta_\mu^*$)  
\begin{equation}
  D_W=\frac{1}{2} \left[ \gamma_\mu (\Delta_\mu^* + \Delta_\mu) - 
  a \, r \, \Delta_\mu^*\Delta_\mu \right]\,, \quad
  \Delta_\mu \psi(x)=\frac{1}{a} \left[ U_{x,\mu} \psi(x+a\hat\mu)- \psi(x) \right]\,.
\end{equation} 
$g$ is the bare coupling, $a$ the lattice spacing and
$F^{\rm clover}_{\mu\nu}$  the standard ``clover-leaf'' form of the
lattice field strength
\begin{equation} 
  F^{\rm clover}_{\mu\nu}=\frac{1}{8 {\rm i} g a^2} \sum_{\mu,\nu=\pm} 
  \left(  U_{x,\mu\nu}^\Box - U_{x,\mu\nu}^{\Box\dagger}\right) \,.
\end{equation}

In the perturbative calculation the operators to be investigated are
sandwiched between off-shell quark states with 4-momenta $p$ and $p'$.
Our calculations are performed in a general covariant gauge, the final numbers
are presented in Feynman gauge for the Wilson parameter $r=1$ and the overlap 
parameter $\rho=1.4$, leaving the coupling strength of the improvement term
$\csw$ free.

In~\cite{Gockeler:2004xb} a detailed discussion of the 
renormalisation procedure for the
case of non-vanishing momentum transfer can be found.
 The matrix of
renormalisation and mixing coefficients $Z_{ik}(a\mu)$ relating
regularised lattice vertex functions $\Gamma_k^L (p',p,a,\gR)$
and $\MS$ renormalised vertex functions $\Gamma_i^R (p',p,\mu,\gR)$
is defined such that
\begin{equation}
  \Gamma_i^R (p',p,\mu,\gR) = Z_\psi^{-1}(a\mu)
  \sum_{k=1}^N Z_{ik}(a\mu) \, \Gamma_k^L (p',p,a,\gR)
  \label{renorm}
\end{equation}
with the quark wave function renormalisation constant
$Z_\psi$. Here $p$ ($p'$) denotes the momentum of the incoming
(outgoing) quark, the renormalisation scale is $\mu$, the
renormalised coupling is denoted by $\gR$, and $N$ is the number
of operators which mix in the one-loop approximation.

\section{Operators and mixing}

We consider quark--antiquark operators with up to two covariant 
symmetric lattice derivatives
$\overset{\leftrightarrow}{D} = \overset{\to}{D}
 - \overset{\leftarrow}{D}$ (for their definitions, see e.g.~\cite{Capitani:2002mp}) 
and external ordinary derivatives $\partial$.

In the non-forward case $q=p'-p\neq 0$ we use two realisations of 
operators with
covariant 
derivatives in momentum space: either the momentum
transfer ``acts'' at the position $x$ associated with the operator (realisation I)
or $q$ is applied at the point half way between the quark fields $\bar{\psi}$ and 
$\psi$ (II). As an example, for two covariant derivatives this leads to
\begin{eqnarray}
  \left(\bar{\psi}\Dlr_\mu\Dlr_\nu \psi\right)^{(II)}(q)&=&
  \frac{1}{a^2}\sum_x \big(\bar{\psi}(x)U_{x,\mu}U_{x+a\hat{\mu},\nu}
   \psi(x+a\hat{\mu}+a\hat{\nu})
  \nonumber\\
  & & \quad -\bar{\psi}(x+a\hat{\nu})U_{x+a\hat{\nu},\mu}
                     U^\dagger_{x+a\hat{\mu},\nu}\psi(x+a\hat{\mu})
  \nonumber\\
  & & \quad -\bar{\psi}(x+a\hat{\mu})U^\dagger_{x,\mu}U_{x,\nu}
                                               \psi(x+a\hat{\nu})
  \nonumber\\
  & & \quad +\bar{\psi}(x+a\hat{\mu}+a\hat{\nu})
    U^\dagger_{x+a\hat{\nu},\mu}U^\dagger_{x,\nu}\psi(x)\big)\,
        \mathrm{e}^{\mathrm{i}q\cdot(x+a\hat{\mu}/2+a\hat{\nu}/2)} \,,
  \label{DDII}
\end{eqnarray}
and realisation I is obtained from (\ref{DDII}) as
\begin{equation}
  \left(\bar{\psi}\Dlr_\mu\Dlr_\nu \psi\right)^{(I)}(q) =
  \cos\left(\frac{aq_\mu}{2}\right)\cos\left(\frac{aq_\nu}{2}\right)
      \left(\bar{\psi}\Dlr_\mu\Dlr_\nu \psi\right)^{(II)}(q) \,.
\end{equation}
We have set the Dirac matrix in the operators equal to the unit
matrix for simplicity.

In what follows, we indicate 
in the operator symbols  the derivatives by superscripts
$D$ and $\partial$.
The quark-antiquark operators with one derivative are given by
\begin{eqnarray}
 \cO^D_{\mu\nu}&=& -\frac{\mathrm i}{2}
  \bar\psi \gamma_\mu \Dlr_\nu \psi\,,
  \ \ \
  \cO^{5,D}_{\mu\nu}= -\frac{\mathrm i}{2}
  \bar\psi \gamma_\mu \gamma_5\Dlr_\nu \psi\,,
  \label{O12} 
\end{eqnarray}
\begin{eqnarray}
  \cO^{T,D}_{\mu\nu\omega}&=&-\frac{\mathrm i}{2}
  \bar\psi [ \gamma_\mu , \gamma_\nu ] \Dlr_\omega \psi \,,
  \ \ \
  \cO^{T,\partial}_{\mu\nu\omega}=-\frac{\mathrm i}{2}
  \partial_\omega
  \left( \bar\psi  [\gamma_\mu,\gamma_\nu ]  \psi \right)\,.
  \label{Oplow2}
\end{eqnarray}
Operators such as $\cO^{T,D}_{\mu\nu\omega}$ 
are of interest for tensor GPDs as well as for transversity and we call them
transversity operators.  For non-chiral fermions, operators
(\ref{Oplow2}) contribute as lower-dimensional operators
to mixing in certain representations which determine second moments of GPDs.

As operators with two derivatives we consider
\begin{eqnarray}
  &&  \cO_{\mu\nu\omega}^{DD} = -\frac{1}{4}
  \bar\psi \gamma_\mu \Dlr_\nu \Dlr_\omega\psi,
  \ \
  \cO_{\mu\nu\omega}^{\partial D}= -\frac{1}{4}
  \partial_\nu \left( \bar\psi \gamma_\mu \Dlr_\omega\psi \right) ,
  \label{OpDD}
  \ \
  \cO_{\mu\nu\omega}^{\partial \partial}= -\frac{1}{4}
  \partial_\nu \partial_\omega
  \left(\bar\psi \gamma_\mu \psi \right)\,,
  \\
  &&  \cO_{\mu\nu\omega}^{5,DD} = -\frac{1}{4}
  \bar\psi \gamma_\mu \gamma_5 \Dlr_\nu \Dlr_\omega\psi,
  \ \
  \cO_{\mu\nu\omega}^{5,\partial D} =  -\frac{1}{4}
  \partial_\nu \left( \bar\psi \gamma_\mu \gamma_5 \Dlr_\omega \psi\right),
  \ \
  \cO_{\mu\nu\omega}^{5,\partial \partial}=-\frac{1}{4}
  \partial_\nu \partial_\omega
  \left(\bar\psi \gamma_\mu \gamma_5\psi \right) 
\nonumber
\end{eqnarray}
and the transversity operators
\begin{equation}
  \cO_{\mu\nu\omega\sigma}^{T,DD}=-\frac{1}{4}\bar\psi [\gamma_\mu,\gamma_\nu]
  \Dlr_\omega   \Dlr_\sigma\psi\,,
  \quad
  \cO_{\mu\nu\omega\sigma}^{T,\partial\partial }=
  -\frac{1}{4} \partial_\omega\partial_\sigma
  \left( \bar\psi [\gamma_\mu,\gamma_\nu] \psi  \right) \,.
  \label{Optrans}
\end{equation}

On the lattice the operators are classified according to the
irreducible representations $\tau_k^{(l)}$ of the hypercubic group
H(4) (for the notation see, e.g., Ref.~\cite{Gockeler:1996mu}).
Here $l$ denotes the dimension of the representation and $k$
labels inequivalent representations of the same dimension. In addition
our operators will be chosen such that they have
definite charge conjugation parity $C$.

To define various index combinations of operators we use the  short-hand notations:
\begin{eqnarray}
  \cO_{\cdots\{ \nu_1 \nu_2 \} }&=&
    \frac{1}{2} \left( \cO_{\cdots \nu_1\nu_2}+\cO_{\cdots \nu_2\nu_1} \right)
  \,,
  \nonumber
  \\
  \cO_{ \{ \nu_1\nu_2\nu_3 \} }&=& \frac{1}{6} \left(
  \cO_{\nu_1\nu_2\nu_3}+\cO_{\nu_1\nu_3\nu_2}+\cO_{\nu_2\nu_1\nu_3} +
  \cO_{\nu_2\nu_3\nu_1}+\cO_{\nu_3\nu_1\nu_2}+\cO_{\nu_3\nu_2\nu_1}  \right)
  \,,
  \\
  \cO_{\|\nu_1\nu_2\nu_3\| } &=& \cO_{\nu_1\nu_2\nu_3}-\cO_{\nu_1\nu_3\nu_2}+
  \cO_{\nu_3\nu_1\nu_2}-\cO_{\nu_3\nu_2\nu_1}-2\,\cO_{\nu_2\nu_3\nu_1}
  +2\,\cO_{\nu_2\nu_1\nu_3}
  \,,
  \nonumber
  \\
  \cO_{\langle\langle\nu_1\nu_2\nu_3\rangle\rangle } &=&
  \cO_{\nu_1\nu_2\nu_3}+\cO_{\nu_1\nu_3\nu_2}
  -\cO_{\nu_3\nu_1\nu_2}-\cO_{\nu_3\nu_2\nu_1}
  \,.
  \nonumber
\end{eqnarray}

For the first moments we choose the following representations and operators:
\begin{equation}
\begin{tabular}{lcc}
  Operator  & Representation & \mbox{$C$} \\
  \hline\\[-2ex]
  $\cO^D_{\{14\}}$
    & $\tau_3^{(6)}$ & $+1$\\ [0.7ex]
  $\cO^D_{44}-\frac{1}{3}\left(\cO^D_{11}+\cO^D_{22}+\cO^D_{33}\right) $
    & $\tau_1^{(3)}$ & $+1$\\ [0.7ex]
  $\cO^{5,D}_{\{14\}}$
    & $\tau_4^{(6)}$ & $-1$\\ [0.7ex]
  $\cO^{5,D}_{44}-\frac{1}{3}\left(\cO^{5,D}_{11}+\cO^{5,D}_{22}
                                           +\cO^{5,D}_{33}\right) $
    & $\tau_4^{(3)}$ & $-1$\\ [0.7ex]
  $\cO^{T,D}_{\langle\langle124\rangle\rangle}$
    & $\tau_2^{(8)}$ & $+1$\\ [0.7ex]
  $\cO^{T,D}_{\langle\langle122\rangle\rangle}-
                         \cO^{T,D}_{\langle\langle133\rangle\rangle}$
    & $\tau_1^{(8)}$ & $+1$\\ [0.7ex]
 \end{tabular}
\vspace*{0.5cm}
\label{tabO}
\end{equation}
All operators in (\ref{tabO}) are multiplicatively renormalisable.
These representations exhaust all possibilities for the twist-2 sector
in the continuum. 

In these proceedings we present results for the
following sets of mixing twist-2 operators 
which are related to the second moments of GPDs in the unpolarised case:
\begin{eqnarray}
  &&\underline{\tau_2^{(4)},\ C=-1}: \ \ \
  \cO_{\{124\}}^{DD} \,, \, \cO_{\{124\}}^{\partial\partial} 
  \label{mixing1}
  \\
  &&\underline{\tau_1^{(8)}, \ C=-1} : 
  \nonumber
  \\
  &&\cO_1=\cO^{DD}_{\{114\}}-\frac{1}{2}
  \left(\cO^{DD}_{\{224\}}+\cO^{DD}_{\{334\}}\right)
  \,, \
  \cO_2=\cO^{\partial\partial}_{\{114\}}-\frac{1}{2} \left(
   \cO^{\partial\partial}_{\{224\}}
  +\cO^{\partial\partial}_{\{334\}}\right)
  \,,
  \nonumber
  \\
  &&\cO_3=\cO^{DD}_{\langle\langle 114\rangle\rangle}-\frac{1}{2}
  \left( \cO^{DD}_{\langle\langle224\rangle\rangle}+
         \cO^{DD}_{\langle\langle334\rangle\rangle}\right)
  \,, \
  \cO_4=\cO^{\partial\partial}_{\langle\langle 114 \rangle\rangle}
  -\frac{1}{2}
     \left(  \cO^{\partial\partial}_{\langle\langle 224 \rangle\rangle}+
          \cO^{\partial\partial}_{\langle\langle 334 \rangle\rangle}\right)
  \,,
  \label{mixing2}
  \\
  &&\cO_5=\cO^{5,\partial D}_{||213||}
  \,, \
  \cO_6=\cO^{5,\partial D}_{\langle\langle213\rangle\rangle}
  \,, \
  \cO_7=\cO^{5,DD}_{||213||}
  \,, \
  \cO_8=   \cO^{T,\partial}_{411}-
  \frac{1}{2}\left(\cO^{T,\partial}_{422}+\cO^{T,\partial}_{433} \right)
  \,.
  \nonumber
\end{eqnarray}
In the polarised case we consider
\begin{eqnarray}
  && \underline{\tau_3^{(4)},C=+1}: \ \ \
  \cO_{\{124\}}^{5,DD} \, , \, \cO_{\{124\}}^{5,\partial\partial} 
  \label{mixing3}
  \\
  && \underline{\tau_2^{(8)},C=+1}:
  \nonumber 
  \\
  &&\cO^5_1=\cO^{5,DD}_{\{114\}}-\frac{1}{2}
  \left(\cO^{5,DD}_{\{224\}}+\cO^{5,DD}_{\{334\}}\right)
  \,,
  \
  \cO^5_2=\cO^{5,\partial\partial}_{\{114\}}-\frac{1}{2} \left(
  \cO^{5,\partial\partial}_{\{224\}}
  +\cO^{5,\partial\partial}_{\{334\}}\right)
  \,,
  \nonumber
  \\
  &&\cO^5_3=\cO^{5,DD}_{\langle\langle 114\rangle\rangle}-\frac{1}{2}
  \left( \cO^{5,DD}_{\langle\langle224\rangle\rangle}+
         \cO^{5,DD}_{\langle\langle334\rangle\rangle}\right)
  \,,
  \
  \cO^5_4=\cO^{5,\partial\partial}_{\langle\langle 114
  \rangle\rangle}-\frac{1}{2}
  \left(  \cO^{5,\partial\partial}_{\langle\langle 224 \rangle\rangle}+
       \cO^{5,\partial\partial}_{\langle\langle 334 \rangle\rangle}\right)
  \,,
  \label{mixing4}
  \\
  &&\cO^5_5=\cO^{\partial D}_{||213||}
  \,,
  \
  \cO^5_6=\cO^{\partial D}_{\langle\langle213\rangle\rangle}
  \,,
  \
  \cO^5_7=\cO^{DD}_{||213||}
  \,,
  \
  \cO^5_8 =
  \cO^{T,D}_{123} -  2\cO^{T,D}_{231} - \cO^{T,D}_{132} \,.
  \nonumber
\end{eqnarray}


\section{Results}

We calculate the matrix elements in one-loop
lattice perturbation theory in the infinite volume limit
following Kawai et al.~\cite{Kawai:1980ja}.
The renormalisation factors for operators related to the first moments 
of GPDs are the same as in the forward case and known for the popular representations.

In the case of the second moments of GPDs and distribution amplitudes  
we present the renormalisation matrices in the form
\begin{eqnarray}
  Z_{ik} (a \mu) &=& \delta_{ik} - \frac{ g_R^2 C_F}{16 \pi^2} 
  \left(\gamma_{ik}\ln (a^2\mu^2) + B_{ik}  \right) 
\end{eqnarray}
where $\gamma$ denotes the matrix of anomalous dimensions and $B$ is 
the matrix of finite one-loop contributions. Below, the numbers in boldface indicate 
results known from the forward case. 

In the representation (\ref{mixing1}) two operators mix in the non-forward case leading to  
\begin{eqnarray}
  \gamma  = \left( \begin{array}{cc}
                      \bf{\frac{25}{6}} & -\frac{5}{6}\\
                           0            &    0
                   \end{array} 
            \right) \,,
	 \  \  \  \
B^{\rm overlap}=\left( \begin{array}{cc}
              {\bf{-47.4441}} & \hspace{-1mm} -0.95719\\
                      0       & \hspace{-3mm}-17.418
                       \end{array}
                \right) \,,
  \label{res1}		
\end{eqnarray}
\begin{eqnarray}
  B^{\rm clover}= \left( \begin{array}{cc}
   {\bf{-11.563}}+ 2.898 \csw -0.984 \csw^2 & \hspace{+3mm} 0.024- 0.255\csw-0.016 \csw^2\\
                      0                     & \hspace{+1mm}20.618+4.746 \csw-0.543 \csw^2
                    \end{array}
         \right) \,.
\end{eqnarray}
The results do not depend on the particular choice of the covariant derivative in 
the non-forward case.

For the representation (\ref{mixing2}) six operators of the same dimension mix 
(in the one-loop approximation the contribution of $\cO_7$ is absent). 
The finite contributions for clover fermions 
$B^{\rm clover}= B^{(0)}+\csw \, B^{(1)}+\csw^2 \, B^{(2)}$ can be found 
in~\cite{Gockeler:2006nb} for the two different derivatives.
For the choice (I) of the derivative we find for overlap fermions
\begin{eqnarray}
\gamma &=& \left(\begin{array}{cccccc}
\bf{\frac{25}{6}} & -\frac{5}{6} & 0           & 0            & 0  & 0\\
     0            &    0         & 0           & 0            & 0& 0\\
     0            &    0         &\bf{\frac{7}{6}} & -\frac{5}{6} & 1 & -\frac{3}{2}\\
     0            &    0         & 0           & 0            & 0& 0\\
     0            &    0         & 0           & 0            & 2& -2\\
     0            &    0         & 0           & 0   & -\frac{2}{3} & \frac{2}{3} \\
                    \end{array}
         \right)
  \label{res21}	 
\end{eqnarray}
\begin{equation}
  B^{\rm overlap}=
    \left( \begin{array}{rrrrrr}
   \bf{-48.1089}       & -3.43393  & \bf{0.53795}       &
          1.33526      &   0.03132 & 0.64459            \\
            0          &  -17.4180 & 0                  &
	    0          & 0         & 0                  \\
  \bf{3.31602 }        &  20.2671  &  \bf{-46.8416}     &
      -10.9597         & -3.52389  & 1.79504            \\
         0             & 0         & 0      & -17.41796   & 0        & 0 \\
         0             & -12.2181  & 0      & 0           & -34.1678 & 1.91865\\
         0             & -12.2181  & 0      & 0           & 0.6396   &  -32.8888
  \end{array}
  \right)
  \label{res22}
\end{equation}

Using clover fermions and the representation (\ref{mixing2}), there is a dangerous 
mixing between the operator $\cO_1$  and the operator $\cO_8$
which is one dimension lower than $\cO_1$:
\begin{eqnarray}
  \cO_1^{\rm clover}\big|_{1/a-{\rm part}} = \frac{g_R^2 C_F}{16 \pi^2}
  (-0.51771+ 0.08325 \, \csw - 0.00983 \, \csw^2)
  {\frac{1}{a}} \cO_8^{{\rm Born}}
\end{eqnarray}
This mixing leads to a contribution which diverges like
the inverse lattice spacing in the continuum limit.
Thus the perturbative calculation of the mixing coefficient is not
reliable and the operator $\cO_8$ has to be subtracted non-perturbatively
from $\cO_1$.
Note, however, that $\cO_1$ and $\cO_8$ are  of {\sl opposite chirality}.

Using chiral lattice fermions in perturbation theory, the general matrix element of 
an operator with two covariant derivatives contains also $1/a$ contributions 
indicating the potential presence of lower dimensional operators at one-loop level.
Choosing a particular irreducible representation such as (\ref{mixing2}), the 
contributing terms are expected to cancel.
This indeed happens in our calculation with overlap fermions and the mixing is 
absent to a very high accuracy numerically:
\begin{eqnarray}
  \cO_1^{\rm overlap}\big|_{1/a-{\rm part}} = 0 \,.
\end{eqnarray}

Using the operators~(\ref{mixing3}) and (\ref{mixing4}) (in the polarised case)
the one-loop results for the chiral overlap fermions coincide with those given 
in (\ref{res1}) and (\ref{res21}),(\ref{res22}), respectively. 

\section{Summary}

We have considered quark-antiquark operators needed for
the computation of the first two moments of GPDs and meson distribution
amplitudes within the framework of lattice QCD. In one-loop lattice
perturbation theory we have calculated the non-forward quark matrix elements
of these operators employing clover improved Wilson fermions and overlap fermions 
together with Wilson's plaquette action for the gauge fields. From the results
we have determined the matrices of renormalisation and mixing
coefficients in the $\MS$-scheme for some 
sets of operators belonging to particular representations
of the hypercubic group.

If there is only mixing between one operator with two
covariant derivatives $D$ and one operator with two external
derivatives $\partial$ (e.g. for the representation (\ref{mixing1}) with three different 
Lorentz indices) the mixing coefficient turns out to be rather small. 
In the case of the representation (\ref{mixing2}) with two equal Lorentz indices 
-- which contains eight potentially mixing operators -- the mixing effects are more
severe. 
As expected, the dangerous mixing with a lower-dimensional operator of opposite chirality
in the case of clover or Wilson fermions
is absent using overlap fermions.
This obviously favours the use of chiral fermions in future simulations related
to GPDs and DAs.

Results for a more complete set of representations using overlap fermions will be presented 
elsewhere~\cite{future}.

\section*{Acknowledgements}
This work
is supported by DFG under contract FOR 465 (Forschergruppe
Gitter-Hadronen-Ph\"{a}nomeno\-logie) and by the EU
Integrated Infrastructure Initiative Hadron Physics under contract
number RII3-CT-2004-506078.
We would like to thank A. Sch\"afer who contributed to part of the results
presented here.

\end{document}